\documentclass[usenatbib]{mn2e}

\usepackage{amsmath}
\usepackage{natbib}
\usepackage{epsfig}
\usepackage{txfonts}

\newcommand{\id}{{\,\rm d}}

\newcommand{\beq}{\begin{equation}}   %

\newcommand{\eeq}{\end{equation}}   %

\newcommand{\beqa}{\begin{eqnarray}}   %

\newcommand{\eeqa}{\end{eqnarray}}   %

\newcommand{\beal}{\begin{align}}

\newcommand{\enal}{\end{align}}

\newcommand{\bspl}{\begin{split}}

\newcommand{\espl}{\end{split}}

\newcommand{\bsub}{\begin{subequations}}

\newcommand{\esub}{\end{subequations}}

\newcommand{\bmulti}{\begin{multline}}   %

\newcommand{\beqm}{\begin{mathletters}}   %

\newcommand{\eeqm}{\end{mathletters}}   %

\newcommand{\Te}{T_{\rm e}}

\newcommand{\Tg}{T_{\gamma}}

\newcommand{\Abl}[2]{\frac{{\rm d} #1}{{\rm d} #2}}

\newcommand{\pot}[2]{#1 \times 10^{#2}}

\newcommand{\ion}[2]{{\text{{\sc #1}\,{\sc #2}}}}

\usepackage{color}
\newcommand{\changeREF}[1]{{#1}}
\newcommand{\changeII}[1]{{#1}}

\newcommand{\changeI}[1]{{#1}}

\title[Recombinations to the Rydberg States of Hydrogen]
{Recombinations to the Rydberg States of Hydrogen and Their Effect During the Cosmological Recombination Epoch}

\author[Chluba, Vasil \& Dursi]{J. Chluba$^{1,2}$\thanks{E-mail: jchluba@cita.utoronto.ca}, G.~M. Vasil$^{1}$\thanks{E-mail: vasil@cita.utoronto.ca}   and \changeI{L.~J. Dursi$^{3,1}$}\thanks{E-mail: ljdursi@cita.utoronto.ca}  
\\
$^{1}$ Canadian Institute for Theoretical Astrophysics, 60 St. George Street,
Toronto, ON M5S 3H8, Canada\\
$^{2}$ Max-Planck Institut f\"ur Astrophysik, Karl-Schwarzschild-Str. 1,
D-85740 Garching, Germany\\
$^{3}$ Scinet, University of Toronto, 256 McCaul Street, Toronto, ON M5T
}

\begin{document}

\date{Received 2010 March 24; Accepted 2010 April 28}

\maketitle

\begin{abstract}
In this paper we discuss the effect of recombinations to highly excited states ($n>100$) in hydrogen during the cosmological recombination epoch.
For this purpose, we developed a new ODE solver for the recombination problem, based on an implicit Gear's method. 
This solver allows us to include up to 350 $l$-resolved shells or $\sim 61\,000$ separate levels in the hydrogen model and to solve the recombination problem for one cosmology in $\sim 27$ hours. This is a huge improvement in performance over our previous recombination code, for which a 100-shell computation ($5050$ separate states) already required $\sim 150$ hours on a single processor. 
We show that for 350 shells down to redshift $z\sim 200$ the results for the free electron fraction have practically converged. 
The final modification in the free electron fraction at $z\sim 200$ decreases from about $\Delta N_{\rm e}/N_{\rm e}\sim 2.8\%$ for 100 shells to $\Delta N_{\rm e}/N_{\rm e}\sim 1.6\%$ for 350 shells. 
However, the associated changes in the CMB power spectra at large multipoles $l$ are rather small, so that for accurate computations in connection with the analysis of {\sc Planck} data already $\sim 100$ shells are expected to be sufficient. 
Nevertheless, the total value of $\tau$ could still be affected at a significant level. 
We also briefly investigate the effect of collisions on the recombination dynamics.  
With our current estimates for the collisional rates we find a correction of \changeREF{$\Delta N_{\rm e}/N_{\rm e}\sim -\pot{8.8}{-4}$ at $z\sim 700$}, which is mainly caused by $l$-changing collisions with protons.
Furthermore, we present results on the cosmological recombination spectrum, showing that at low frequencies collisional processes are important.
However, the current accuracy of collisional rates is insufficient for precise computations of templates for the recombination spectrum at $\nu \lesssim 1\,$GHz, and also the effect of collisions on the recombination dynamics suffers from the uncertainty in these rates. 
Improvements of collisional rates will therefore become necessary in order to obtain a final answer regarding their effects during recombination.
\end{abstract}

\begin{keywords}
Cosmic Microwave Background: cosmological recombination, temperature
  anisotropies, cosmological recombination spectrum, spectral distortions
\end{keywords}

\section{Introduction}
\label{sec:Intro}
Close to the maximum of the Thomson visibility function \citep{Sunyaev1970} at redshift $z\sim 1100$, and slightly before, the dynamics of hydrogen recombination is mainly controlled by the net 2s-1s two-photon decay rate and the escape of photons from the Lyman-$\alpha$ resonance \citep{Zeldovich68, Peebles68, Sunyaev2009}.
These two channels are the main 'bottlenecks' during the epoch of hydrogen recombination, with the Lyman-$\alpha$ channel being more important at high redshifts ($z\gtrsim 1300-1400$) and the 2s-1s two-photon decay channel dominating at lower redshifts \citep[e.g. see][]{Jose2006}.
In total about 57\% of all hydrogen atoms became neutral via the 2s-1s two-photon decay channel, while some 43\% of the hydrogen electrons recombined through the Lyman-$\alpha$ channel \citep{Chluba2006b}.

At redshift $z\sim 1100$, the rate at which electrons meet a proton and recombine is still very large, even though only a much smaller fraction of these recombinations actually end with an electron settling into the ground state.
\changeI{Under such circumstances the {\it exact} rate at which electrons are captured by protons is not absolutely crucial for precise computations of the cosmological ionization history, as long as this rate is close enough to the true value, and does not lead to an artificial 'bottleneck' caused by the {\it incompleteness} of the used atomic model for hydrogen.}

At lower redshifts ($z\lesssim 800-900$), however, the rate of recombinations drops significantly, because ever fewer free electrons and protons are available, so that the number densities of free electrons and protons start\footnote{For conditions in our Universe, the free electron and proton number densities never freeze out completely, but at low redshifts they enter a period over which they evolve very slowly.} to 'freeze out'.
During and just before this period, the total electron capture rate becomes one of the additional 'bottlenecks' of cosmological recombination, and the {\it precise} value of the total recombination cross section starts to become very important.
Under these circumstances the {\it completeness} of the used atomic model for hydrogen becomes one of the key elements for accurate computations of the cosmological recombination history, which aim at reaching a level of precision down to $\sim 0.1\%$.

With measurements of the cosmic microwave background (CMB) temperature and polarization power spectra, as currently carried out with the {\sc Planck} Surveyor, it indeed has become very important to understand the ionization history at this level of precision \citep[e.g. see][]{Hu1995,Seljak2003, Lewis2006}.
In particular, our ability to determine the precise value of the primordial spectral index of scalar fluctuations, $n_{\rm s}$, one of the key parameters to learn more about inflation \citep[e.g. see][for recent constraints]{Komatsu2010}, could be severely compromised if the detailed physics of cosmological recombination are not understood \citep[e.g.][for recent discussion]{Jose2010}.
Over the past few year, this fact has motivated a large number of works on the physics of recombination \citep[e.g. see][]{Dubrovich2005, Chluba2006, Kholu2006, Switzer2007I, Wong2007, Jose2008, Karshenboim2008, Hirata2008, Chluba2008a, Jentschura2009, Labzowsky2009, Grin2009}, all with the aim to get ready for the analysis of CMB data from {\sc Planck}, {\sc Act}, {\sc Spt} and in the future from {\sc Cmbpol}.

In this paper we want to discuss the effect of recombinations to highly excited states ($n>100$) in hydrogen during cosmological recombination ($z\sim 1000$) on both the cosmological recombination history and the recombination spectrum.
In particular, we want to focus on the effects associated with the detailed evolution of the populations in the angular momentum sub-states of hydrogen, a process that already has received some attention earlier \citep{Jose2006, Chluba2007}.
However, the previous computations were limited to models with 100 $l$-resolved shells in hydrogen \citep{Chluba2007}, amounting to a total of $5050$ levels.
As pointed out there, in order to obtain converged results for the free electron fraction at low redshifts one has to include $\sim 200-300$ shells, a task that requires refined numerical methods and significant improvements of the earlier recombination code.

Recently, \citet{Grin2009} strongly advanced such computations, including up to 250 shells, or a total of $\sim 31\,000$ separate states, during hydrogen recombination. 
They conclude, that for the computations of the CMB temperature and polarization power spectra the results for the cosmological recombination history are converged at the $\sim 0.5\sigma$ at Fisher-matrix level when including 128 $l$-resolved shells in the atomic model of hydrogen. 
Here we also reach a similar conclusion, using a completely different numerical method and independent recombination code, showing that for the {\sc Planck} data analysis $\sim 100$ shells should already be sufficient at large multipoles $l$.
Nevertheless, due to huge improvements in the performance of our recombination code, it is now possible to solve a recombination history for 100 shells in about 11 min on 8 cores, and even 350-shell computations only take a little longer than a day.
Therefore, it will be easy to account for this correction providing a new training set for \changeREF{the multi-dimensional regression code {\sc Rico}} \citep{Fendt2009}.

We also present results for the cosmological recombination spectrum from hydrogen \citep{Dubrovich1975, DubroVlad95, Kholu2005, Jose2006, Chluba2006b, Sunyaev2009} and provide more detailed computations which include the effect of collisional processes.
In particular, we develop a new solver for the coupled system of ordinary differential equations (ODEs), which can be easily adapted to other problems, e.g. for computations of nuclear networks appearing in supernova and star formation calculations \citep[e.g. see][]{Timmes1999, Hix2006}, or for chemical networks which are important during reionization \citep[e.g. see][and references therein]{Shapiro1987, Anninos1997, Tegmark1997, Abel1997, Gnedin2009} and the dark ages \citep[e.g. see][]{Stancil1996, Stancil1998, Schleicher2008}.

The paper is structured as follows: in Sect.~\ref{sec:problem} we give a few details and references about the cosmological recombination problem. For more basic overview we refer the interested reader to \citet{Seager2000} and \citet{Sunyaev2009}.
In Sect.~\ref{sec:algorithm} we provide some details about the new recombination code and the ODE solver that we developed. This section is rather technical, and only meant for the interested reader.
In Sect.~\ref{sec:results} we discuss the results for the cosmological recombination history and the recombination spectrum. There we also briefly discuss the effects of collisions during recombination, and conclude in Sect.~\ref{sec:conc}.

\section{Description of the recombination problem}
\label{sec:problem}
Although the current version of our recombination code \citep{Chluba2009c} allows us to include several additional processes that already have been shown to be important to precise computations of the recombination history \citep{Fendt2009, Jose2010}, here we want to focus on the effect of recombinations to highly excited level ($n>100$) in hydrogen, taking into account the detailed evolution of the populations in the angular momentum levels in each shell with principle quantum number $n$.
We therefore restrict ourselves to a minimal model of hydrogen and helium, that does not include any of the detailed radiative transfer effects, such as photon feedback \citep[][]{Chluba2007b, Switzer2007I, Kholupenko2009, Chluba2009c} or Lyman-$\alpha$ diffusion \citep{Hirata2009, Chluba2009b}.
Details about our atomic model for hydrogen can be found in \citet{Jose2006} and \citet{Chluba2007}. The helium model is explained in \citet{Jose2008} and \citet{Chluba2009c}.
Details about the setup of the rate equations can be found in \citet{Seager2000}.

However, for the computations carried out in this paper, we do include the acceleration of helium recombination caused by the absorption of resonant photons by hydrogen  \citep{Kholupenko2007, Switzer2007I, Jose2008}.
This process allows helium to finish recombining until redshift $z\sim 1700$, so that accounting for this process is very important for the initial condition of the hydrogen recombination problem, which at $z\sim 1650$ can be approximated using the Saha-equations.
In most of our computations we use this setup, however, when computing the cosmological recombination spectrum we evolve both hydrogen and helium starting at $z=3400$ with Saha-values for their populations.

\subsection{Numerical computations of the recombination and photoionization rates}
\label{sec:Ric_Rci}
One of the crucial and also rather time-consuming aspects of the recombination problem is the \changeREF{separate} computation of the photoionization and recombination rates \changeREF{for each considered level}.
Here several points are important. For high levels the $l$-dependence of the photoionization cross section, \changeREF{$\sigma_{i\rm c}(\nu)$ for level $i=(n,l)$}, is very strong. This implies that for large $n$ both recombination and photoionization are mainly occurring through the low-$l$ states, as the Gaunt-factors drop very strongly with $l$ \citep[e.g. see][]{Chluba2007}. 
Also the energy-dependence \changeREF{of the photoionization cross section becomes exponential for states with $n\gg1$ and $l \sim n$}, so that care must be taken when integrating the cross sections over the CMB blackbody spectrum.

\changeREF{In addition, for high $n$ states\footnote{\changeREF{It is straightforward to show that in our Universe stimulated recombinations become $\sim 10$ times stronger than spontaneous recombinations for levels with principal quantum number $n\gtrsim 23\,\left[\frac{1+z}{1100}\right]^{-1/2}$.}} the effect of stimulated recombinations caused by the presence of CMB blackbody photons has to be included.
As shown earlier  \citep{Chluba2007}, not only the recombination rate for each level increases due to stimulated recombinations, but also the $l$-dependence of the recombination coefficients within a highly excited shell is strongly affected.}

Furthermore, for large $n$ \changeREF{we encountered} difficulties with the recursion relations for the photoionization cross section given by \citet{StoreyHum1991}, when going far away (several hundred times) from the threshold frequency $\nu_{i\rm c}$. In this case the initial condition for the recursion relation numerically became zero, so that the cross section is effectively zero for all $l$ states.
We avoid this problem by rescaling the terms in the recursion formulae with the $n^{\rm th}$ root of the initial condition, which one could easily compute. With this the computations of the cross sections becomes stable up to very large $n$ ($\sim 1000$) and $l$, and very large distances from $\nu_{i\rm c}$.

In order to choose the range of integration most efficiently, we identified the frequency $\nu_{\rm s}>\nu_{i\rm c}$ at which $\sigma_{i\rm c}\nu^2$ \changeREF{becomes} extremely small\footnote{\changeREF{Since the integrand of the photonionization coefficient, $\beta_{i\rm c}$, in the CMB blackbody field scales as $\sigma_{i\rm c}\nu^2/[e^{h\nu/k T_\gamma}-1]\lesssim (k T_\gamma/h)\, \sigma_{i\rm c}\nu \propto \nu^{-2}$, while $\sigma_{i\rm c}\nu^2 \propto \nu^{-1}$,  the latter provides a rather conservative estimate for the behaviour of the integrand with frequency. Note that only the ratio to the value at the threshold frequency is important, so that $\sigma_{i\rm c}\nu^2$ drops slower towards larger $\nu$ than $\sigma_{i\rm c}\nu$.}} ($\sim 10^{-30}-10^{-40}$ of the value at the threshold). 
For high $n$ and $l$ states, this occurred over only a few threshold energies, while for the low $l$ states we \changeREF{found a much wider range}.
\changeREF{
However, we typically limited the frequency interval to $\nu_{i\rm c}\lesssim \nu \lesssim 10^8\,\nu_{i\rm c}$, and in the Wien tail of the CMB blackbody spectrum we in addition utilized the exponential cutoff for $h\nu \gg k T_\gamma$.}

\changeREF{Also}, to avoid the time-consuming recursive computation of the cross sections at every time-step, we tabulated them on a dense grid and then use spline interpolation. This procedure accelerates \changeREF{the} computations of the recombination and photoionization rates by a large factor, without significant loss of accuracy.

To carry out the recombination integrals we implemented a fully adaptive integrator based on the integration formulae given by \citet{Patterson1968}. 
This method is based on Gaussian quadrature rules, but in contrast to simpler Gaussian rules (e.g. Gauss-Kronrad) these formulae are fully nested, so that function evaluations of {\it every} previous subdivision can be reused.
We also attempted a scheme based on Chebychev integration rules, but eventually the formulae by \citet{Patterson1968} performed better.
We compared our results for the recombination and photoionization rates with those obtained using {\sc Nag}-integrators, and found excellent agreement (to the level of precision which we used for the integration; this was normally $\sim 10^{-8}-10^{-9}$ in relative terms).

\section{The new multi-level recombination code}
\label{sec:algorithm}
Based on our previous multi-level hydrogen and helium recombination code \citep{Chluba2009c}, we developed a new solver for the system of coupled ordinary differential equations. In particular, we replaced all routines that were used from the commercial {\sc Nag} library\footnote{See http://www.nag.co.uk/numeric/}, so that our new code now is completely non-commercial, and parts of it could be parallelized.

For the development of this new solver on a single processor machine three points turned out to be very important: 
(i) because of the vastly different timescales involved in the evolution of the populations of the excited states in hydrogen and helium, we use a stiffly stable algorithms with adaptive step-size control; 
(ii) because of the size of the Jacobian matrix for the equation system it is important to make use of its sparseness for both, {\it storage} and {\it matrix operations};
(iii) it is also important to use analytic expressions for the Jacobian matrix when possible, in order to achieve sufficient numerical stability and accuracy, and also to make the code faster.

After managing these aspects we also parallelized parts of our code using OpenMP. However the largest boost in our performance (a factor $\sim 150$ in comparison to our previous version) is obtained because of the improvements in connection with (ii) and (iii).
Below we provide some more details on each point.

\subsection{The stiffly stable ODE solver}
\label{sec:stiffness}
The system of rate equations describing the cosmological recombination problem is very {\it stiff}. Even in the effective three-level approximation, which is used in the implementation of {\sc Recfast} \citep{SeagerRecfast1999}, because of the vastly different timescales on which the electron temperature and the population of the hydrogen ground state adjust their values, the stiffness of the equations is already so large that a stiffly stable solver is required to achieve high accuracy and good performance.

Below we explain the essential parts of the ODE solver we developed for the recombination problem. 
This solver was implemented in a general way, so that it also can be applied to other problems, as already pointed out in the introduction.
We would also like to mention, that in comparison to the commercial {\sc Nag} sparse ODE solver routine {D02NJF} we were able to achieve a slightly better performance. This is likely due to the higher order of the implicit equations for the time-stepping (6th order instead of 5th), the efficient use of sparseness in the equation setup, and the direct and complete ability to tune all solver parameters independently.
Furthermore, with the {\sc Nag} solver we were unable to solve the recombination problem for more than 150 shells.

\subsubsection{Gear's method with adaptive time-step}
For stiff problems implicit methods are known to be numerically more desirable than explicit methods, because of both performance and stability issues \citep[e.g. see][]{NUMRES}.
We chose an implicit Gear's method \citep{Gear1971} to evolve the solution to the recombination problem from one time $t_i$ to the next $t_{i+1}$. 
This method is related to the backward difference formulae (BDF) methods \citep[see][]{Curtiss1952}. 

The differential equation system for the recombination problem can be cast into the form
\beal
\label{eq:dy_dt_ODE}
\Abl{y}{t}&=f(t, y),
\end{align}
where $y=y(t)$ is the (high-dimensional) solution vector of the problem at time $t$. 
Denoting the solution at time $t_i$ as $y_i$ the implicit equation for the time-step using the Gear's method reads
\beal
\label{eq:dy_dt_Gear}
y_{i+1}&=h\,\lambda\,\dot{y}_{i+1}+ \sum_{k=0}^{5} \kappa_k\,y_{i-k},
\end{align}
with $\dot{y}_{i+1}=\id y_{i+1}/\id t=f(t_{i+1}, y_{i+1})$, $h=t_{i+1}-t_i$, and $y_{i-k}$ being the solution to the recombination problem at previous times $t_{i-k}<t_{i}$.

It was shown by Gear in 1971 that only up to 6th order such equation is stiffly stable \citep[see][]{Gear1971}, so that we truncated the sum in Eq.~\eqref{eq:dy_dt_Gear} at $k=5$.
The coefficients $\lambda$ and $\kappa_i$ can be determined using the Taylor-expansion of the solution $y_{i+1}$. Since initially only the solution at $t_0$ is known, and because we wanted to allow both changing order and variable time-step we re-derived these coefficients up to sixth order. 
They can be found in Appendix \ref{app:Gears_coeffies}. For constant time-step they resemble those given in \citet{Antia}.

\subsubsection{Solving the non-linear equation system}
Equation \eqref{eq:dy_dt_Gear} defines a system of non-linear equations that determines the solution $y_{i+1}$ at time $t_{i+1}$.
One can linearize this equation assuming that one has a guess for the solution $y^{p}_{i+1}$ which is {\it close} to the correct one.
Inserting $y^{p}_{i+1}$ on the right hand side of Eq.~\eqref{eq:dy_dt_Gear} one obtains $\tilde{y}^{p}_{i+1}$.
This then leads to
\beal
\label{eq:LA_Gear}
[{\bf 1}-h\,\beta\,{\bf J}_f ]\,\delta y^{p}_{i+1}&=\tilde{y}^{p}_{i+1}-y^{p}_{i+1},
\end{align}
where ${\bf J}_f$ is the Jacobian matrix of the system Eq.~\eqref{eq:dy_dt_ODE} evaluated at $y=y^{p}_{i+1}$, and \changeREF{$\delta y^p_{i+1}=y^{p+1}_{i+1}-y^{p}_{i+1}$, where $y^{p+1}_{i+1}$ is the solution to Eq.~\eqref{eq:LA_Gear}}. 
Note that in general $\delta y^p_{i+1}\neq \tilde{y}^{p}_{i+1}-y^{p}_{i+1}$.
After obtaining $\delta y^p_{i+1}$ from Eq.~\eqref{eq:LA_Gear} one can use $y^{p+1}_{i+1}=y^{p}_{i+1}+\delta y^p_{i+1}$ as new initial condition for the problem and then iterate until convergence is reached. We usually use a combination of relative and absolute error control to check for convergence.
Especially for the free electron fraction, the electron temperature, and the number of electrons in the ground states of hydrogen and helium, \changeREF{tight settings  ($\epsilon \sim 10^{-8}$) for the relative accuracies were necessary}.

To solve for $\delta y^{p}_{i+1}$ it is not useful to explicitly invert the matrix ${\bf A}=[{\bf 1}-h\,\beta\,{\bf J}_f]$. Even though ${\bf 1}-h\,\beta\,{\bf J}_f $ might be very sparse, the inverse $[{\bf 1}-h\,\beta\,{\bf J}_f ]^{-1}$ likely will be dense. 
\changeI{Also, inversion will only lead to numerically meaningful results for well-conditioned matrices.}
In addition, for the recombination problem the sparseness is of order percent (see Sect.~\ref{sec:Jacobian}), so that simple multiplications $b={\bf A}\, x$ are fast, while one expects a very large number of operations to invert $[{\bf 1}-h\,\beta\,{\bf J}_f ]$. 

We therefore choose an iterative scheme, based on a stabilized bi-conjugate gradient method \citep[e.g. see][]{Barrett1993}.
This method generates two orthogonal sequences of vectors for the matrix $\bf A$. These vector sequences are the residuals of the iterations, which are iterated until convergence is reached.
As preconditioning we simply use the inverse diagonal elements of the Jacobian. For the recombination problem this is sufficient.

In order to avoid many {\it expensive} evaluations of the Jacobian matrix, we furthermore make use of Broyden's method \citep{Broyden1965, Gag1979, NUMRES}.
Here we assume that in each update only the non-zero matrix elements should be changed, so that the sparseness of the matrix is preserved during the whole run.
We find that this approximation works very well for the recombination problem.

To obtain an initial guess for the solution at each time-step we use a simple extrapolation method based on the solution at previous time-steps.
This allows us to write
\beal
\label{eq:dy_dt_Extra}
y_{i+1}&= \sum_{k=0}^{5} \gamma_k\,y_{i-k},
\end{align}
where the \changeREF{coefficients} $\gamma_i$ are also given in Appendix~\ref{app:Gears_coeffies}.

\begin{figure}
\centering
\includegraphics[width=0.85\columnwidth]{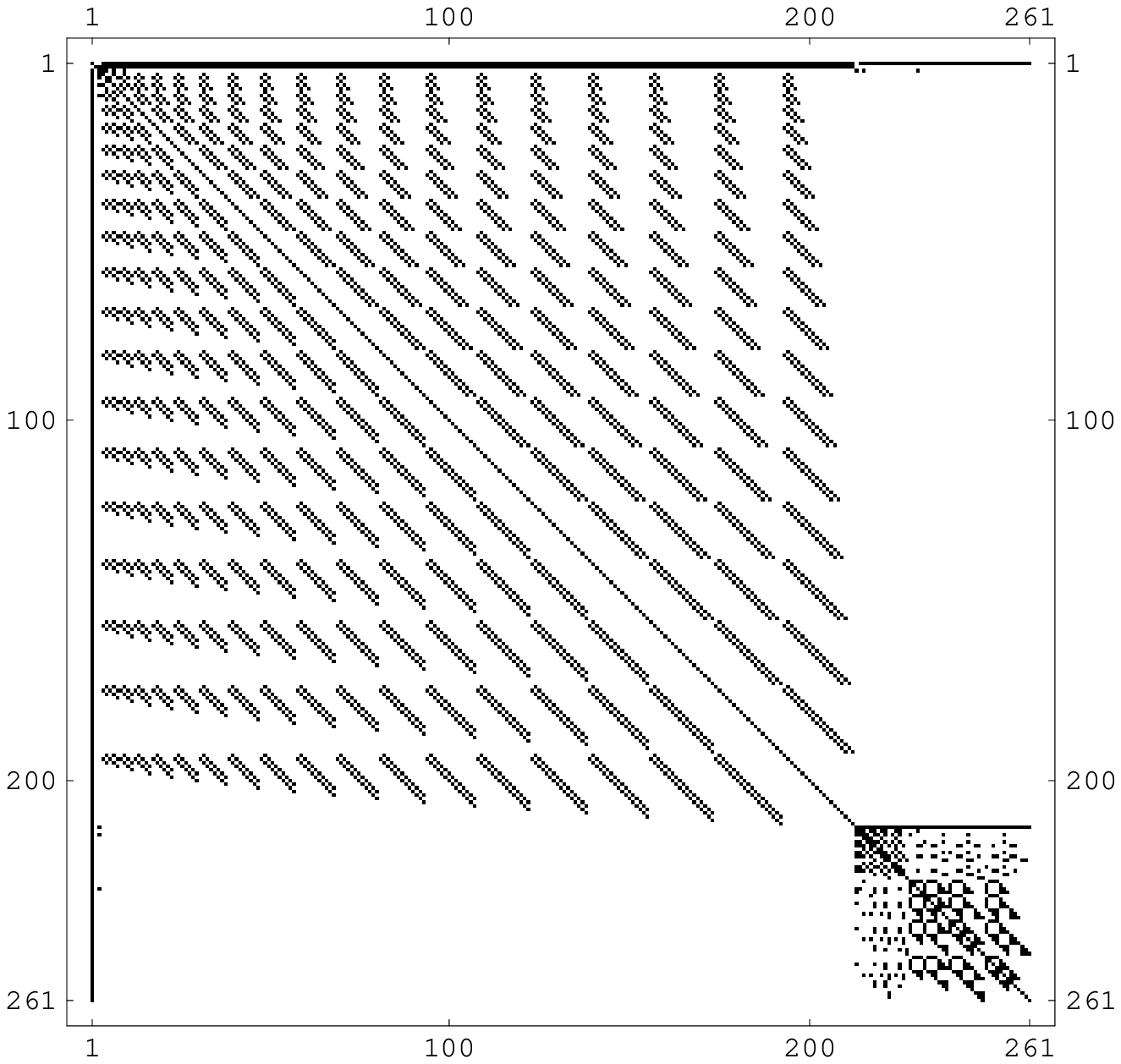}
\\
\includegraphics[width=0.85\columnwidth]{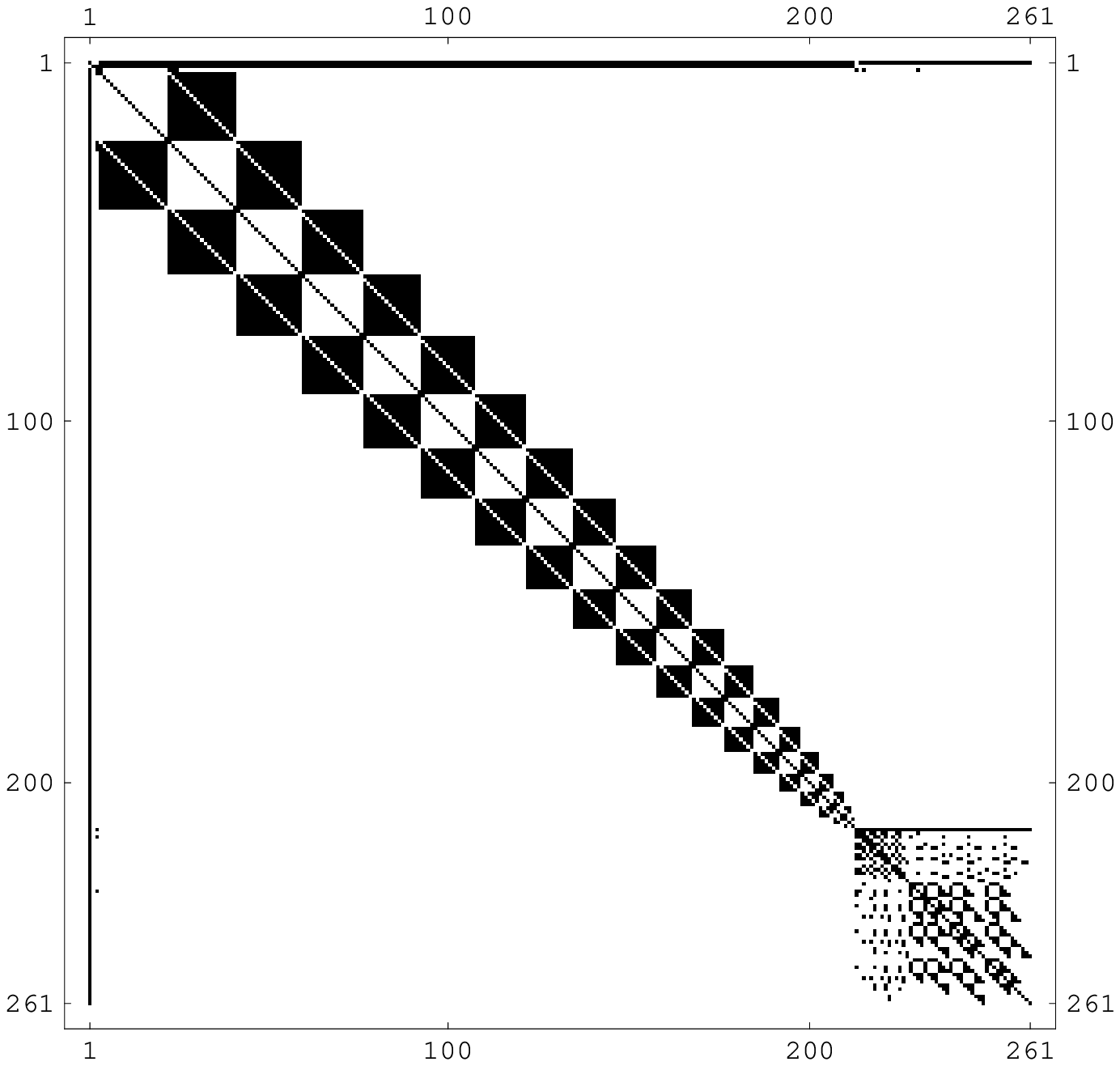}
\\
\includegraphics[width=0.85\columnwidth]{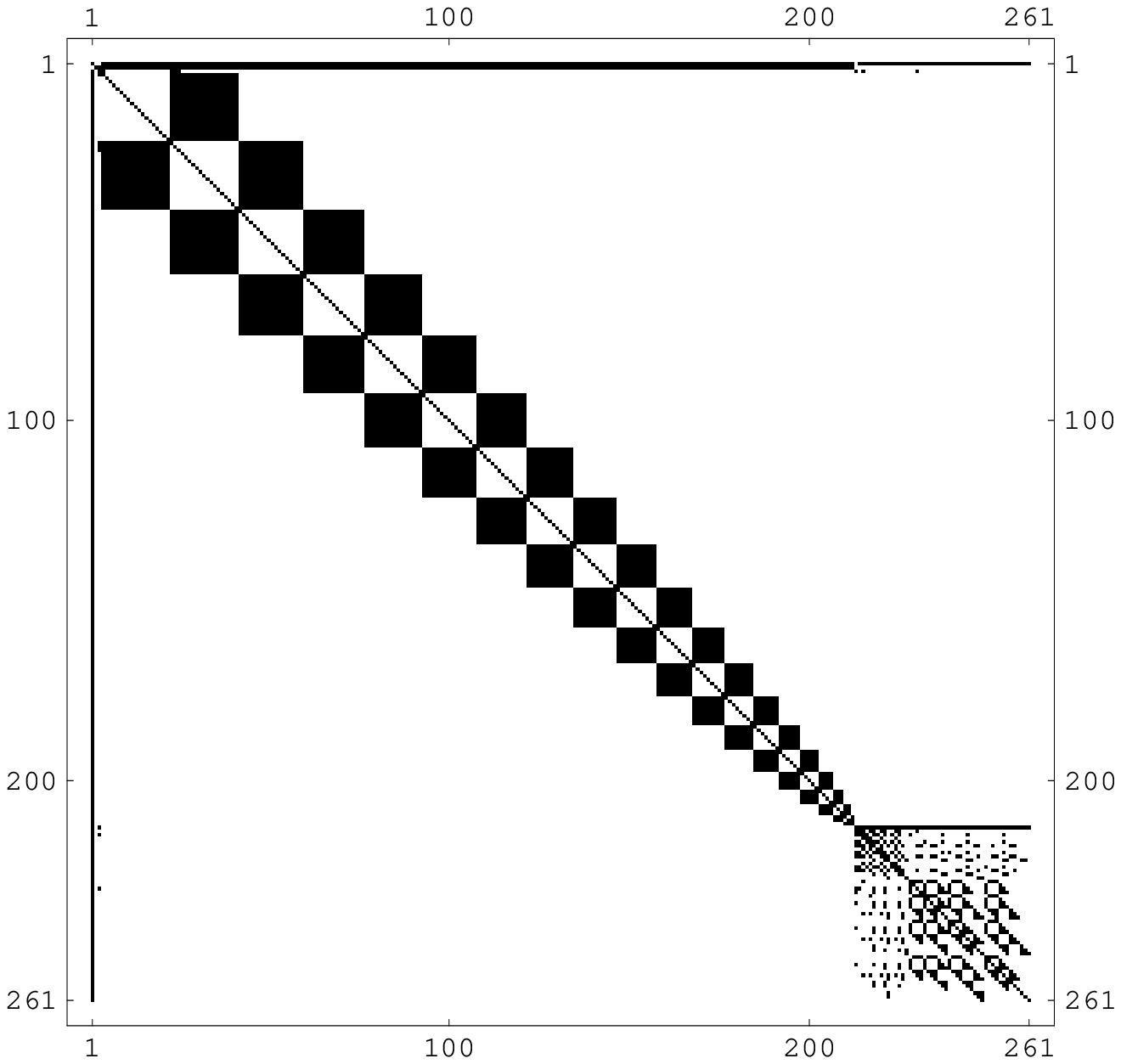}
\caption{Structure of the Jacobian matrix for different cases: 
top panel -- ordering A; middle panel -- ordering B; lower panel -- ordering B with $l$-changing collisions. 
%
In all cases 20 shells for hydrogen and 5 shells for helium were included. The equations related to helium mainly lead to entries in the lower right corner of the Jacobian. The sparseness is about $10\%$.}
\label{fig:Jacobian}
\end{figure}
\subsubsection{Ordering of the equations}
We tried several orderings of the equations, however, none of them lead to significant improvements of the performance. In particular, we also tried the ordering of the hydrogen states suggested by \citet{Grin2009}, without any apparent benefit. In their ordering, states with the same angular quantum number $l$ are grouped and then ordered in the sequence $l=0, 1, 2, ...\, n_{\rm max}-1$, where $n_{\rm max}$ denotes the largest shell that was included. In our final implementation we use
\beal
\label{eq:sol_ordering}
y(t)&=
\left(\begin{array}{c}
X_{\rm e}\\
\Te\\
X^{\rm HI}_{i}\\
X^{\rm He I}_i\\
\end{array}
\right),
\end{align}
where the populations of the hydrogen states, $X^{\rm HI}_{i}$, are ordered like 1s, 2s, 2p, 3s, 3p, 3d, 4s, 4p, etc. and the helium states, $X^{\rm He I}_i$, are ordered in the same manner, first for the singlet and subsequently for the triplet atom. 
This is the same ordering used in earlier implementations of the problem \citep{Jose2006, Chluba2007}. 
The distribution of elements in the Jacobian matrix for ordering A \citep[as in][]{Jose2006, Chluba2007}, and ordering B \citep[like in][]{Grin2009} are illustrated in Fig.~\ref{fig:Jacobian}.

\subsubsection{Quasi-stationary approximation for the excited states}
The excited states ($n>1$) adjust their populations on time-scales much shorter than the expansion time. It is therefore possible to simplify the differential equation system using the quasi-stationary approximation for levels with $n>1$.
We tried such approach, but found that in that case the performance of our solver was reduced. 

We also tried assuming that only for levels with $n\gg1$ the quasi-stationary approximation was applicable, again with no gain in performance.
We therefore used the full system of ODE's in our computations.
However, we would like to note that even in the quasi-stationary approximation for levels with $n>1$ we could obtain accurate recombination histories and also recombination spectra.
For the recombination problem the quasi-stationary approximation seem valid for the excited states, however, the performance of our ODE solver was not {suggesting} such an approach.

Roughly speaking, the quasi-stationary approximation amounts simply to dropping the ${\bf 1}$ term in Eq.~\eqref{eq:LA_Gear} relative to the  $h\,\beta\,{\bf J}_f$ and re-normalizing with respect to $h$.  When considered in this way, it is clear that the majority of the effort already results from solving a large system of non-linear equations.  Furthermore, leaving the ${\bf 1}$ term amounts to a physical regularization should the Jacobian become formally singular.

\subsection{Sparseness of the Jacobian matrix and its setup}
\label{sec:Jacobian}
The size of the Jacobian matrix for the ODE system scales like $\sim n^4/4$ with the total number of shells, $n$, that are included into the atomic model of hydrogen.
For example, the recombination model for a 100-shell hydrogen atom with resolved angular momentum quantum numbers leads to a system of $\sim 5050$ differential equations, and hence a Jocobian with $\sim 25$ million entries.
For a 350-shell hydrogen model one already deals with about $\sim 61\,000$ ODEs and the full Jacobian has $\sim 3.7$ billion entries.

However, because of the dipole selection rules, the transition matrix is rather sparse. Simple estimates show that the number of non-zero elements scales $\sim \frac{2}{3} n^3$, so that the matrix is sparse at a level of percent for 100 shells\footnote{For $\sim 270$ shells the sparseness drops below one percent.}. 
\changeREF{One can find} examples for the structure of the Jacobian in Fig.~\ref{fig:Jacobian}.

\changeI{Because of the scaling of the number of non-zero elements with $n$ the Jacobian becomes even more sparse for larger $n$.}
In computations, it therefore is important to make use of this sparseness, in both {\it storage} and {\it matrix operations}.
For this purpose, we adapted the {\sc SparseLib$++$} Library\footnote{See http://math.nist.gov/sparselib++/} to store the Jacobian matrix.
For the purpose of parallelization we used the compressed-column format.
Also, the required linear algebra operations were implemented using this library.
Here only operations with non-zero elements were performed, so that the efficiency is very large as compared to full matrix routines.

To compute the Jacobian of the system Eq.~\eqref{eq:dy_dt_ODE}, we use both analytic and numerical derivatives. For all the bound-bound dipole transitions of hydrogen, we use fully analytic expressions, while for helium at this stage we always use numerical derivatives. These are computed with a two-point central difference formula, which is second-order accurate in the chosen\footnote{
\changeI{We also tried higher order formulae, but the tradeoff caused by the additional number of function evaluations was too large. With the second-order formula we normally used $\delta X\sim 10^{-6} \,X$.}}  $\delta X$.
Also we compute the derivative of the recombination rates with respect to $\Te$ analytically, leading to an additional recombination integral in the matrix setup.
A typical evaluation of the Jacobian for $\sim 100$ shells in our current implementation requires about $\sim 6$ seconds on a standard single-processor machine. With \changeI{parallelization using OpenMP} we were able to gain a factor of $3-4$ for this evaluation on 8 cores.

We also comment, that we usually solved for the free electron fraction using $X_{\rm e}=1-\sum X^{\rm HI}_i+f_{\rm HeI}-\sum X^{\rm H}_i$, instead of the differential equation. This is possible due to particle conservation. 
In our current solver we simply replaced the ODE for the free electron fraction by the above algebraic equation.

\begin{table}
\caption{Performance of the recombination code. $n_{\rm max}$ denotes the included number of hydrogen shells, and $n_{\rm eq}$ gives the number of hydrogen levels. 'Run A' is executed on a standard single-processor machine (MacBook Pro, 2.4 GHz Intel Core 2 Duo, 3 GB 667 MHz DDR2 SDRAM), while for 'Run A (S $m$)' and 'Run B (S $m$)'  \changeI{we used nodes of the SciNet GPC supercomputer, each with two 2.53 GHz quad-core Intel Nehalem E5540s, and 16GB of 1066 MHz DDR3 SDRAM,} where $m$ gives the number of cores that were used. For additional details on the parameters for the different runs see Sect.~\ref{sec:Perform}.}
\label{tab:perform}
\centering
\begin{tabular}{@{}cccccc}
\hline
\hline
$n_{\rm max}$ & $n_{\rm eq}$ & Run A & Run A (S 1) & Run A (S 8) & Run B (S 8) \\
\hline
100 & 5050 & 52 min & 33 min & 11 min & 25 min \\
150 & 11325 & 3.4 h & 2.0 h & 1.3 h & 2.8 h \\
200 & 20100 & 9.7 h & 5.5 h & 3.6 h & 8.0 h \\
250 & 31375 & 22 h & 12 h & 8.0 h & 17 h \\
300 & 45150 & --  & 23 h & 14 h & 35 h \\
350 & 61425 & -- & 42 h & 27 h & 63 h \\
\hline
\hline
\end{tabular}
\end{table}
\subsection{Performance of the code}
\label{sec:Perform}
The performance of our new recombination code initially is mainly limited by the speed of the matrix vector operation ${\bf A}\, x$ and the setup of the recombination and photoionization rates. 

To reduce the effort in connection with the recombination integrals, we use the tabulation scheme described in \citet{Chluba2008c}. We again confirm the precision of this procedure, by comparing with computations in which all the recombination rates were explicitly evaluated at each time-step.
Also we tabulate the changes in the escape probabilities of the \ion{He}{i} $2^{1}{\rm P}_1-1^{1}{\rm S}_0$ and $2^{3}{\rm P}_1-1^{1}{\rm S}_0$ resonances caused by the presence of neutral hydrogen \citep{Kholupenko2007, Switzer2007I, Jose2008}, using the results of a calculation for a 5-shell hydrogen with 5-shell helium model.
With this setup, for a 100-shell-hydrogen atom about $40\%$ of all time is spent for multiplications ${\bf A}\, x$, while about $\sim 20\%$ of time is spent computing the recombination integrals. For a larger number of shells the contribution from the multiplications ${\bf A}\, x$ strongly increases (e.g. reaching $\sim 70\%$ for 200 shells), so that parallelization of this part is beneficial.
However, by parallelization we currently only achieve an additional factor of $\sim 2$ in comparison to a single-processor machine.

To measure the performance of our code we compared runs with different settings for the accuracy. If one is only interested in the cosmological recombination history, it in principle is possible to run the code in a faster mode. 
The helium recombination history can already be computed rather precisely with 5 shells for hydrogen and 5 shells for helium. It is then possible to start the computation of hydrogen recombination just after helium has become neutral, including a much larger amount of hydrogen levels.
For the spectral distortions from hydrogen on the other hand it is important to evolve the whole system starting well before helium recombination, since even hydrogen is emitting some amount of photons in every transition because of the reprocessing of helium photons.

In Table~\ref{tab:perform} we show a few examples regarding the performance of our code. For Run A we use a simplified helium recombination history, and only start the full hydrogen recombination calculation at $z=1650$ evolving everything until $z=200$. 
For Run B we start at redshift $z=3400$ and evolve both hydrogen and helium until $z=200$.
Due to memory restrictions we did not compute cases with more than 250 shells on a single-processor machine.

As can be seen from Table~\ref{tab:perform}, for Run A and  $n_{\rm max}\gtrsim 200$ on a single-processor machine the computational time scales roughly $\propto n^{3.3-3.6}_{\rm max}$ with the total number of hydrogen shells, or like $\propto n^{1.7-1.8}_{\rm eq}$ with the total number of equations.
Starting the computation at redshift $z=3400$ instead of $1650$ take about twice as long.
On a single core our code seems to be a factor of $\sim 10$ faster\footnote{Their run for 200 shells hydrogen starting at redshift $z\sim 1606$ takes about 4 days on a single-processor machine..} than the one of \citet{Grin2009}.
%

\changeI{On SciNet's GPC system\footnote{See http://www.scinet.utoronto.ca/}, we find an immediate performance gain of a factor of approximately 1.6 on a single processor, despite the very small change in clock speed; this is almost certainly due to the improved memory bandwidth of the newer Nehalem processors.   The SparseLib++ matrix operations were parallelized with OpenMP; the matrix-vector multiply was decomposed by counting the number of matrix non-zeros and forcing a static decomposition which split the number of non-zeros approximately equally over the number of threads; other matrix operations were parallelized with dynamic partitioning over rows.}
\changeI{In addition, we parallelized the Jacobian matrix setup, including the computations of the recombination and photoionization integrals needed for updates of the interpolation tables. This allowed} us to gain another factor of $\sim 2$ using 8 cores, so that eventually a computation of  the recombination history and recombination spectrum was about $2-3$ \changeREF{times} faster than on a standard single-processor machine.
For Run B on 8 cores and for $n_{\rm max}\gtrsim 200$ the computational time again scaled roughly $\propto n^{3.4-3.5}_{\rm max}$ with the total number of hydrogen shells, or like $\propto n^{1.7-1.8}_{\rm eq}$ with the total number of equations.

We would like to mention that one could further speed the computations of the recombination history up when ignoring accuracy for the cosmological recombination spectrum. With this one would likely be able to gain another factor $\sim 2-3$. For example, we always limited our step-size to $\Delta z\lesssim 3$, to obtain a resolved representation of the recombination spectrum. This could easily be increased twice. Another possibility is to allow {\it overshooting} beyond the desired redshift point, and then interpolating the solution where it is requested. This will diminish the accuracy for the spectrum, but still leads to precise results for the recombination history. 
However, at this point we did not follow this idea any further.

\section{Results}
\label{sec:results}
In this section we present our results for the cosmological recombination history and the cosmological recombination spectrum. We first discuss the effect of the completeness of the atomic model and then present some first results in connections with collisional processes, which we incorporate using simple approximations.

\begin{figure}
\centering
\includegraphics[width=0.95\columnwidth]{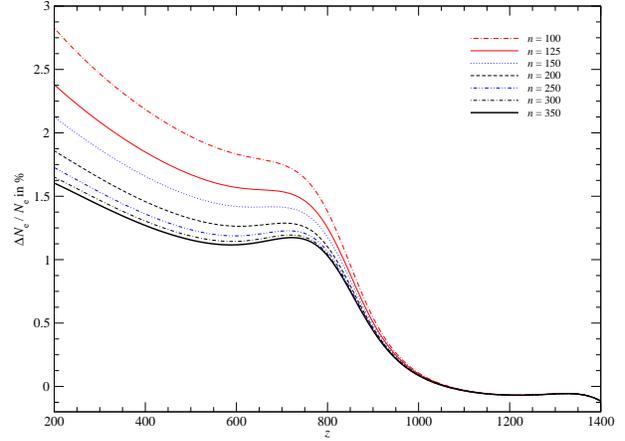}
\caption{Correction to the ionization history for different number of hydrogen shells as a function of redshift. \changeREF{Shown is the relative difference with respect to {\sc Recfast} (see text for details)}. One can see that at high redshifts ($z\gtrsim 900$) the dependence on $n_{\rm max}$ is very small, while at low redshifts the results seem to start converging for 350 shells.}
\label{fig:Xe}
\end{figure}

\begin{figure}
\centering
\includegraphics[width=0.95\columnwidth]{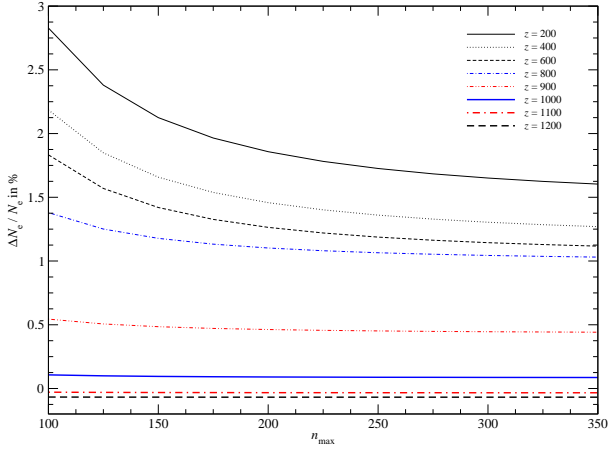}
\caption{Correction to the ionization history at different redshifts as a function of the the total number of hydrogen shells.}
\label{fig:DXe_n}
\end{figure}

\begin{figure}
\centering
\includegraphics[width=0.95\columnwidth]{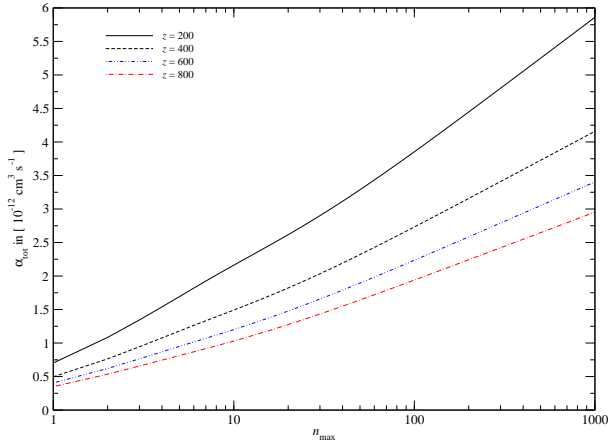}
\caption{Total recombination coefficient, $\alpha_{\rm tot}=\sum_{n,l} \alpha_{nl}$, as a function of $n_{\rm max}$ and for different redshifts. For simplicity we assumed that $\Te=\Tg$, and we also included the effect of stimulated recombinations in the ambient CMB blackbody field.}
\label{fig:alpha}
\end{figure}

\subsection{The cosmological recombination history}
\label{sec:Xe}
In this section we present the results for the corrections to the cosmological recombination history. We compared with the {\sc Recfast~v1.4.2} code \citep{Wong2008}, but excluding the corrections to the helium recombination history and getting rid of the switches in the ODE system \citep[see][for details]{Fendt2009}.
\changeREF{We used the standard hydrogen fudge factor, $f_{\rm H}=1.14$.} 
In Fig.~\ref{fig:Xe} we present the results for the modification of the free electron fraction as a function of redshift including a different number of hydrogen shells \changeREF{(see Sect.~\ref{sec:RECFASTUPDATE} for a more detailed discussion on the shape of the correction)}.
One can clearly see that at high redshifts, close to the maximum of the Thomson visibility function ($z\gtrsim 1000$), the dependence of the correction on the number of hydrogen shells is already rather small.
On the other hand at low redshifts, it is important to include shells up to $n_{\rm max }\sim 300-350$ to obtain fully converged results for the ionization history. 

This can be seen even more clearly in Fig.~\ref{fig:DXe_n}, which shows the dependence of the correction at fixed redshifts $z$ on $n_{\rm max}$. 
Apparently, at $z\sim 200$ the change to the recombination history starts to converge for $n_{\rm max}\sim 300-350$, while at $z\gtrsim 900$ already $n_{\rm max}\sim 100$ is sufficient.
As mentioned in the introduction, this is because at $z\gtrsim 800-900$ the dynamics of recombination is strongly controlled by the escape of photons from the Lyman-$\alpha$ resonance and the net two-photon decay-rate of the 2s level, so that increasing the effective recombination rate cannot affect the ionization history very much. 
This was already pointed out in \citet{Fendt2009}, in connection with the effect of the {\sc Recfast} hydrogen fudge factor on the recombination history.

On the other hand, at low redshift, the completeness of the atomic model is still important, and the 'bottle-neck' for recombination is not only set by the 2s-1s transition rate or the Lyman-$\alpha$ channel, but also by the capture rate of electrons from the continuum.
However, at the level of percent to the correction at $z\sim 200$ the atomic model seems to become complete for $n_{\rm max}\sim 350$.

%
We would like to note that the simple total recombination coefficient, $\alpha_{\rm tot}=\sum_{n,l} \alpha_{nl}$, continues to grow for $n_{\rm max}>350$ (see Fig.~\ref{fig:alpha}). Although $\alpha_{\rm tot}$ does not capture any of the dynamical effects (e.g. net radiative transition rates and escape probabilities; possible collisional ionizations for very high levels) that define the effective recombination coefficient, Fig.~\ref{fig:alpha} shows that in principle it is possible to increase the effective recombination rate, when the bottle-necks of recombination set by the effective rate of transitions to the ground state will be modified. 
Such kind of modification could for example be achieved by collisions that mix different $l$ sub-states, as we will discuss in Sect.~\ref{sec:collisions}.

We would also like to mention that until now we have not checked the convergence of the results at lower redshifts ($z\lesssim 200$). There the chemistry of the Universe will also become important. In addition, large uncertainties in the reionization physics will affect our ability to carry out precise computations.
Resolving these questions is beyond the scope of this paper.

\begin{figure}
\centering
\includegraphics[width=0.95\columnwidth]{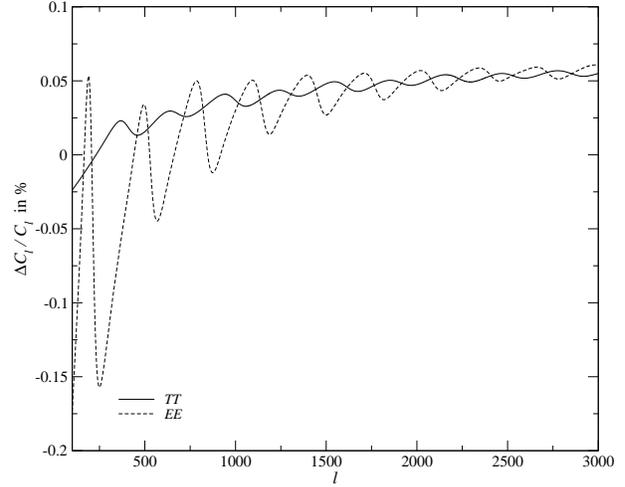}
\caption{Changes to the CMB temperature and polarization power spectra when going from 100 shells to 350 shells. The curves were obtained using a modified version of {\sc Cmbeasy} \citep{Doran2005}.}
\label{fig:DCl}
\end{figure}
\subsubsection{Effect on the CMB power spectra}
\label{sec:Cls}
For the analysis of {\sc Planck} data, only the changes in the CMB temperature and polarization power spectra really matter.
In Fig.~\ref{fig:DCl} we show the correction that is obtained when going from 100 to 350 shells.
This correction is very small, only reaching $\Delta C_l/C_l \sim 0.05\%$ at $l=3000$, and also remaining well below the cosmic variance limit at small $l$'s. 
Therefore, one does not expect any important changes to the biases in $n_{\rm s}$ and $\Omega_{\rm b} h^2$ obtained earlier by \citet{Jose2010} using the results for a 110-shell hydrogen models. 
Regarding the {\sc Planck} data analysis $\sim 100$ shells already seem to be sufficient.
\changeREF{Nevertheless, due to the large improvement in the performance of our recombination code, it will be rather easy to account for the full correction, e.g. providing an updated training set for {\sc Rico} \citep{Fendt2009}}

Although the final correction to the free electron fraction at $z\sim 200$ decreases by a factor $\sim 1.8$ when going from 100 to 350 shells (see Fig.~\ref{fig:Xe}), the modification in the CMB power spectra at large multipoles $l$ is much smaller, simply because the main effect is connected with a change in the total optical depth $\tau$, while the position of the maximum ($z\sim 1100$) and the width ($\Delta z\sim 200$) of the visibility function remain practically unchanged. 
However, the value of $\tau$ \changeREF{could still be} affected at the level of percent by this modification. One therefore expects a change in the $EE$ power spectrum at low multipoles $l$, close to the region that is also affected by reionization \citep[e.g. see][and references therein]{Haiman2003, Colombo2009a}. We indeed found a $\Delta C_l/C_l \sim -0.8\%$ correction to the $EE$ power spectrum at low multipoles ($l\lesssim 10$), while the $TT$ power spectrum remained nearly unaltered. 
However, our computations did not include any detailed reionization model, so that it is difficult to estimate the impact of this correction regarding the value of $\tau$.
Similarly, one will have to include possible effects of dark matter annihilation \citep{Padmanabhan2005, Zhang2006, Galli2009, Slatyer2009, Cirelli2009, Huetsi2009, Kanzaki2009} or decaying particles \citep{Chen2004, Zhang2007} into such considerations.
This is beyond the scope of this paper.

\subsubsection{\changeREF{Possible approaches for improvements to {\sc Recfast}}}
\label{sec:RECFASTUPDATE}
\begin{figure}
\centering
\includegraphics[width=0.95\columnwidth]{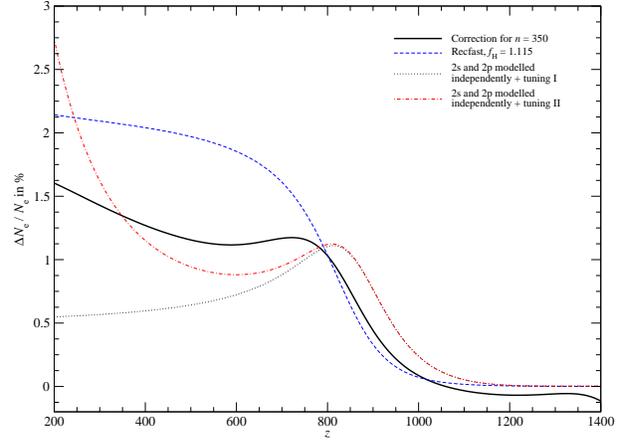}
\caption{\changeREF{Correction to the ionization history with respect to {\sc Recfast} for different recombination models. The solid/black curve shows the result from our multi-level recombination code for $n_{\rm max}=350$.
The other curves were computed using {\sc Recfast}, with different modifications relative to the standard case (see text for details)}.}
\label{fig:IMPROVE}
\end{figure}

\changeREF{The shape of the correction to the free electron fraction for our 350 shell computation (cf. Fig.~\ref{fig:Xe} and also the solid/black line in Fig.~\ref{fig:IMPROVE}), suggests that in comparison with {\sc Recfast} there are (at least) two different regimes: the region at $z\gtrsim 800$ and the low redshift region at $z\lesssim 800$. 
At $z\gtrsim 800$ the correction grows with a larger slope than at $z\lesssim 800$. Also the correction shows some small local maximum at $z\sim 720$ and a local minimum at $z\sim 590$.
What are the likely physical reasons for this behaviour?
}

\changeREF{
The are several aspects of the recombination problem that currently cannot be captured by {\sc Recfast}.
First one might think about the effective recombination coefficient\footnote{\changeREF{Here the subscript 'B' stand for case B recombination, where direct recombinations to the ground state are excluded \citep[e.g. see][]{Baker1938, Hummer1994}.}}, $\alpha_{\rm B}$, which in {\sc Recfast} is probably slightly overestimated.  
There,  $\alpha_{\rm B}$ is modelled using the fitting formulae given by \citet{Pequignot1991} plus some additional overall hydrogen fudge factor $f_{\rm H}=1.14$ \citep{SeagerRecfast1999}.
By adjusting the hydrogen fudge factor one can compensate part of the correction, as already pointed out earlier \citep[e.g. see][]{Fendt2009}.
Here we found that for the 350 shell results $f_{\rm H}=1.110-1.115$ captures the amplitude and rough scaling of the modification around $z\sim 900$, but overestimates the effect at $z\lesssim 800$ by a factor of $1.5-2.0$ (see Fig.~\ref{fig:IMPROVE}, blue/dashed curve).
Also, by simply changing $f_{\rm H}$ it is impossible to reproduce the local maximum or minimum at $z\sim 720$ and $z\sim 590$.
}

\changeREF{
There are two additional aspects of the recombination problem that change at $z\sim 800$: the temperature of the matter starts to depart significantly from the photon temperature. 
This will modify details of the interplay between the photon field with the hydrogen atom, so that $\alpha_{\rm B}$ becomes a function of both electron and photon temperature. This aspect cannot be captured with the simple formulae given by \citet{Pequignot1991}, and without additional computations it is hard to estimate the importance of this effect. 
However, as we show below, the temperature scaling of $\alpha_{\rm B}$ indeed changes the low redshift behaviour of the correction. 
} 

\changeREF{
In addition, around the same redshift the 2s-1s  two-photon channel again becomes less important than the Lyman-$\alpha$ transition, since the optical depth in the Lyman-$\alpha$ resonance drops. 
This has more delicate consequences, which currently cannot be captured by {\sc Recfast}. Since in the derivation of the {\sc Recfast} equations it is assumed that the 2s and 2p state are in full statistical equilibrium ($N_{2p}=3 \,N_{\rm 2s}$), every electron that reaches either the 2s or the 2p state is immediately shared with the other.
It was already shown earlier, that this assumption is not valid at the end of recombination, since collisional processes are inefficient \citep{Chluba2007}.
Avoiding this approximation therefore would allow to distinguish between electrons that reach the 2s state and the 2p state.
A simple derivation (using the quasi-stationary assumption for the evolution of the 2s and 2p populations) yields a slightly modified {\sc Recfast} equation for hydrogen
\beal
\label{eq:NEW_RECFAST}
\Abl{X^{\rm H}_{\rm e}}{z}&=\frac{1}{H(z) (1+z)}\,
\left\{ 
C_{\rm 2s}\left[ X_{\rm e} N_{\rm p}\alpha_{\rm 2s} - X_{\rm 1s}\,\xi \,\beta_{\rm 2s}\right] 
\right.
\nonumber\\
&\qquad\qquad\qquad\qquad
+\left.
C_{\rm 2p}\left[ X_{\rm e} N_{\rm p}\alpha_{\rm 2p} - 3\,X_{\rm 1s}\,\xi \,\beta_{\rm 2p}\right] 
\right\},
\end{align}
where $H(z)$ is the Hubble expansion factor, $X_{\rm e}$ and $X_{\rm 1s}$ are the free electron fraction and 1s-population, $N_{\rm p}=[1-X_{\rm 1s}]\,N_{\rm H}$, and $N_{\rm H}$ is the total number of hydrogen nuclei in the Universe.
Furthermore, the recombination and photonionization rates for the 2s and 2p state are denoted as $\alpha_{i}$ and $\beta_{i}$, respectively. 
Due to the presence of CMB photons one also obtains the factor $\xi=\exp(-h\nu_{21}/k\,T_\gamma)$, where $\nu_{21}\approx \pot{2.47}{15}\,{\rm Hz}$ is the Lyman-$\alpha$ transition frequency, and the 2s and 2p inhibition factors are given by
%
\bsub
\beal
\label{eq:C_factors}
C_{\rm 2s}&=\frac{A_{\rm 2s1s}}{A_{\rm 2s1s}+\beta_{\rm 2s}}
\\
C_{\rm 2p}&=\frac{A^{\ast}_{\rm 2p1s}}{A^{\ast}_{\rm 2p1s}+\beta_{\rm 2p}}.
\end{align}
\esub
Here $A_{\rm 2s1s}$ is the 2s-1s two-photon decay rate, 
and the effective Lyman-$\alpha$ transition rate is given by
$A^{\ast}_{\rm 2p1s}=P_{\rm S} A_{\rm 2p1s}\approx \left(\frac{3\lambda^3 N_{\rm 1s}}{8\pi\,H(z)}\right)^{-1}$, where $P_{\rm S}$ is the Sobolev escape probability of the Lyman-$\alpha$ resonance, and $A_{\rm 2p1s}\approx \pot{6.27}{8}\,{\rm s^{-1}}$ is the Lyman-$\alpha$ transition rate.
}

\changeREF{The important point about Eq.~\eqref{eq:NEW_RECFAST} is that one has to specify the partial recombination rates to the 2s and 2p state. Using detailed balance one then also obtains the partial photonionzation rates just like in {\sc Recfast}.
This allows to account for more details in the microphysics of the multi-level cascade.
Given $\alpha_{\rm B}$, one would normally assume $\alpha_{\rm 2s}\approx \frac{1}{4}\,\alpha_{\rm B}$ and $\alpha_{\rm 2s}\approx \frac{3}{4}\,\alpha_{\rm B}$, so that $\beta_{\rm 2s}\approx \frac{1}{4}\,\beta_{\rm B}$ and $\beta_{\rm 2p}\approx \frac{1}{4}\,\beta_{\rm B}$, where $\beta_{\rm B}$ is the photoionization coefficient in the normal {\sc Recfast} case.
However, the  detailed dynamics of recombination connected with the cascade of electrons from higher levels to the 2s and 2p state (these are included into the computation of $\alpha_{\rm B}$), can render this approximation crude, in particular at the end of recombination.
One aspect that is connected with this, is that the 2s and 2p state will departure from full statistical equilibrium, as seen earlier \citep{Jose2006, Chluba2007}. 
}

\changeREF{To demonstrate the principle possibilities of Eq.~\eqref{eq:NEW_RECFAST}, we show the resulting correction with respect to the normal {\sc Recfast}, choosing $f_{\rm H}=1.135$,  $\alpha_{\rm 2s}= 0.59\,\alpha_{\rm B}$, and $\alpha_{\rm 2s}= 0.41\,\alpha_{\rm B}$ \changeII{(Fig.~\ref{fig:IMPROVE}, curve label with 'tuning I')}. 
As can be seen, there now is a maximum at $z\sim 800$, which is actually produced by the fact that the Lyman-$\alpha$ transitions again starts controlling the recombination process.
Note that in general the ratio $\gamma=\alpha_{\rm 2s}/\alpha_{\rm 2p}$ should be a function of redshift, so that one can accommodate more details of by computing $\alpha_{\rm 2s}$ and $\alpha_{\rm 2p}$ using the results of our full recombination code.
}

\changeREF{Finally, we also computed the correction for  $f_{\rm H}=1.135$,  $\alpha_{\rm 2s}= 0.59\,\alpha_{\rm B}$, and $\alpha_{\rm 2s}= 0.41\,\alpha_{\rm B}$, but using $\alpha_{\rm B}=\alpha_{\rm B}(T_{\gamma})$ instead of $\alpha_{\rm B}=\alpha_{\rm B}(T_{\rm e})$ \changeII{(Fig.~\ref{fig:IMPROVE}, curve label with 'tuning II')}.
Recombinations physically are controlled by both the electron temperature and the photon temperature. The latter enters, because CMB photon strongly control the radiative cascade within the hydrogen atom, and we just wanted to show, that varying the temperature dependence of $\alpha_{\rm B}$ in fact has most effect at low redshifts.
We therefore expect that a detailed computation of $\alpha_{\rm 2s}$ and $\alpha_{\rm 2p}$ from the full results of our recombination code, may allow us to capture more aspects of the microphysics. 
}

\changeREF{
Along the same line, one may consider including more levels into the {\sc Recfast} code (e.g. 3s, 3d, 3p). Again one could determine the effective recombination rates from the results of our recombination code for each included level.
Such approach may allow to capture in detail some of the feedback processes within the Lyman-series \citep{Chluba2007b, Kholupenko2009}. In addition, simple modifications should allow to include the effect of stimulated 2s-1s two-photon decays and the reabsorption of Lyman-$\alpha$ photons in the 1s-2s channel \citep{Chluba2006, Kholu2006}.
In combination with our new ODE solver, such extensions will probably not degrade the speed of {\sc Recfast} very much, so that it can \changeII{still} be used in the analysis of future CMB data.
We plan to investigate these possibilities in the future.
}

\begin{figure}
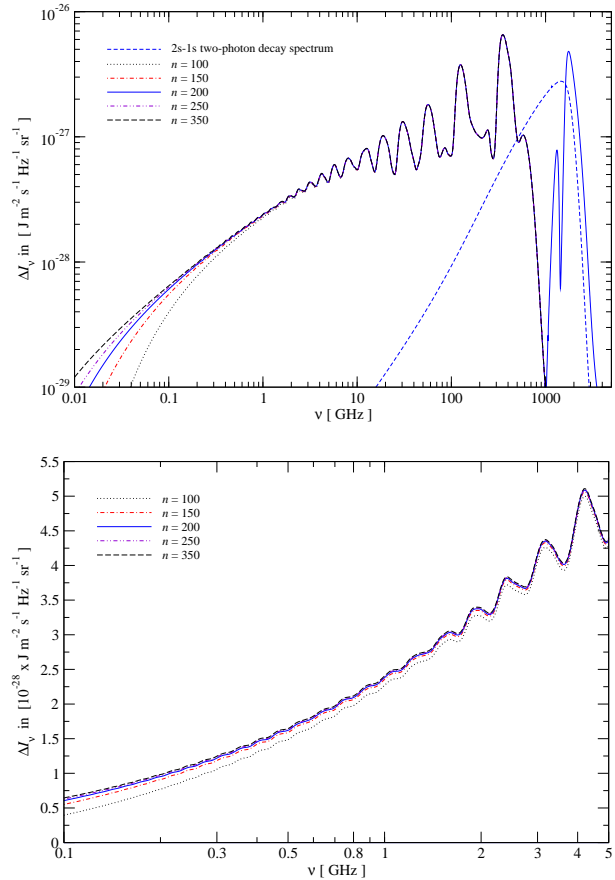

\centering
\includegraphics[width=0.95\columnwidth]{./eps/Spec.eps}
\\[4mm]
\includegraphics[width=0.95\columnwidth]{./eps/Spec.zoom.eps}
\caption{The bound-bound cosmological recombination spectrum of hydrogen. The lower panel shows a zoom in on the frequency range around $\nu\sim 1\,$GHz. Free-free absorption and collisions were not included.}
\label{fig:spec}
\end{figure}
\subsection{The cosmological recombination spectrum of hydrogen at low frequencies}
\label{sec:spec}
Including more than 100 shells in the computation of the hydrogen recombination spectrum should mainly affect the low frequency part of the recombination spectrum \citep{Chluba2007}.
This is because including more shells enables more electrons to pass through highly excited states, emitting additional low frequency photons in the cascade towards lower levels.

In Fig.~\ref{fig:spec} one can observe this effect. The recombination radiation from hydrogen still increases by a factor $\sim 1.6$ at $\nu\sim 0.1\,$GHz, when going from 100 to 350 shells. 
However, at this level the spectrum seems to converge. 
From the observational point of view the region $\nu\gtrsim 1\,$GHz is much more interesting. There it may become possible to measure the variable component of the recombination spectrum in the future \citep[see][for overview]{Sunyaev2009}.
At $\nu\gtrsim 1\,$GHz the recombination spectrum seems to converge at the level of a few percent and better, when including 350 shells.
However, this statement is only true regarding the completeness of the atomic model for hydrogen. 
The effect of $l$-changing collisions still alters the shape of the recombination spectrum at low frequencies \citep{Chluba2007}, as we will further discuss in the next section.
Also at frequencies $\nu\lesssim 0.1\,$GHz free-free absorption should become important \citep{Chluba2007}, so that accurate computations in this region are more involved.

\begin{figure}
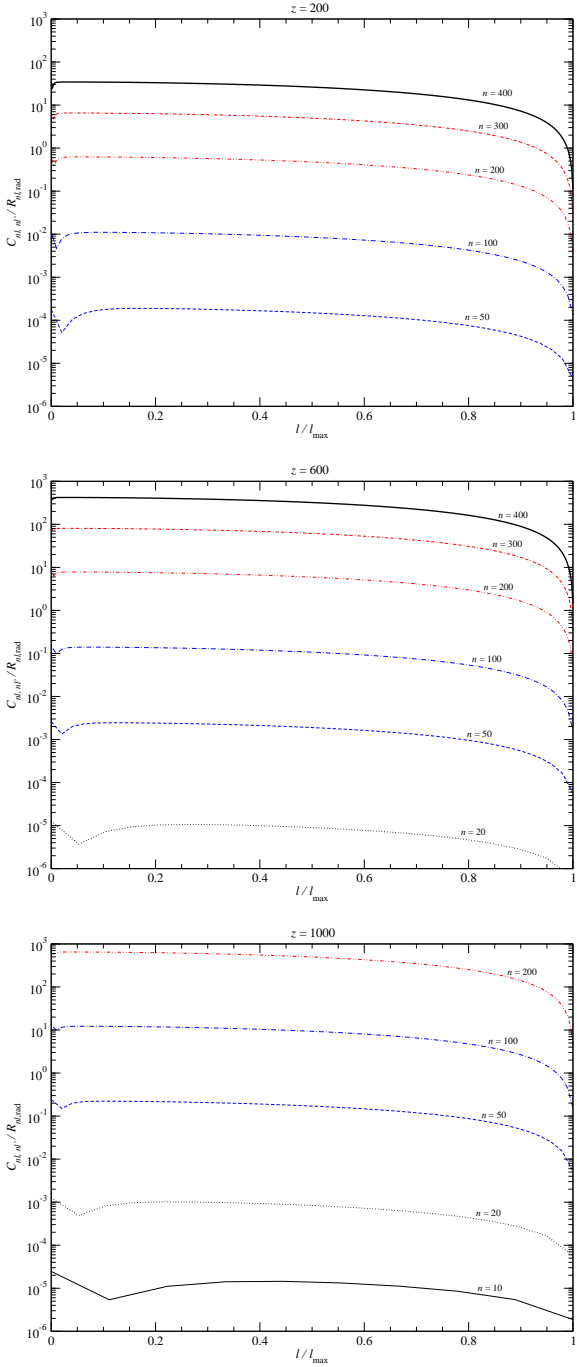

\centering
\includegraphics[width=0.90\columnwidth]{./eps/Cnl_nplp.z_200.eps}
\\[3mm]
\includegraphics[width=0.90\columnwidth]{./eps/Cnl_nplp.z_600.eps}
\\[3mm]
\includegraphics[width=0.90\columnwidth]{./eps/Cnl_nplp.z_1000.eps}
\caption{Ratio of the total $l$-changing collision rate caused by electron and proton impact to the total radiative transition rate in the CMB blackbody radiation field for different shells of hydrogen and at different redshifts as a function of $l$. Here $l_{\rm max}=n-1$.
The kink at low $l$ is due to transitions out of the $n\rm p$ states, which depopulate fastest.}
\label{fig:l-colls}
\end{figure}
\subsection{The effect of collisions}
\label{sec:collisions}
Already in our earlier work \citep{Chluba2007} we discussed the effect of collisions on the cosmological recombination spectrum and the ionization history of our Universe. However, because of computational restriction \changeREF{in that implementation} we were unable to include more than 100 shells for hydrogen. 
\changeREF{However}, it is expected that in a diluted plasma such as in our Universe in the epoch of recombination, the effects of collisions will become more important for higher shells. 
The most important aspect of this problem is that collisions can lead to mixing of $l$ and $n$-states in the Rydberg levels. One expects that $l$-mixing becomes important for $n\gtrsim n_{l-\rm mix}$, and then $n$-mixing and mixing with the continuum starts at $n\gtrsim n_{n-\rm mix}$, where $n_{n-\rm mix}\gg n_{l-\rm mix}\gg 1$, \changeREF{since $n$-changing collisions (which require transfer of energy) are less effective than $l$-changing collision.}

Here we investigate the effect of collisions on both the ionization history and the cosmological recombination spectrum for up to \changeREF{300} shells.
%
%
We include $l$-changing collisions, $n$-changing collisions, and collisional ionizations by electron, proton and $\alpha$-particle (only important at $z\gtrsim 1800$) impact in our computations, as described in \citet{Chluba2007}. 
In that work, the main collisional rates were computed using simple approximations given by \citet{Rege1962, Pengelly1964, Brocklehurst1971}.
For additional overview see also \citet{Mashonkina1996}.

A more detailed treatment of collisional processes is far beyond the scope of this paper, but given that the effects on the low frequency recombination spectrum are significant, it may become important to refine these computations. 
With our current estimates for the collisional rates, the dynamics of recombination do not seem to be affected at a level that is very important for the CMB power spectra, however, we would like to point out that the accuracy of the used collisional rates easily allows for factors of a few.
Therefore, it will be very important to refine the current calculations of the collisional rates, in order to give a final answer for their relevance during recombination.

Additionally, at low redshifts, collisions with neutral hydrogen atoms could start to become important \citep[see][for recent computations of rate coefficients]{Mihajlov2004}. However, we defer this problem to future work.

\begin{figure}
\centering
\includegraphics[width=0.95\columnwidth]{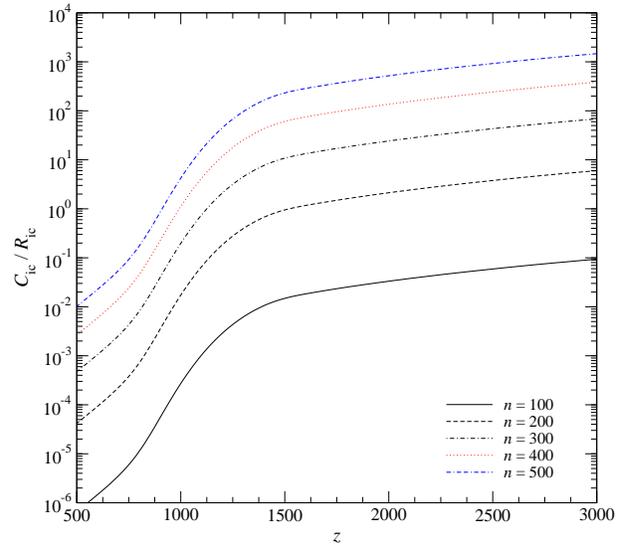}
\caption{Ratio of the total collisional ionization rate induced by electron impact to the total photoionization rate caused by absorption of photons from the CMB blackbody for different shells of hydrogen. At $z\gtrsim 1000$ collisional ionization is always faster than photoionization for $n\gtrsim 400-500$. Note however, that collisional ionization is much slower (by a factor of $\sim 10^2-10^3$) than bound-bound radiative transitions until much larger redshifts ($z\gtrsim 3000-4000$). Collisional ionization therefore never seems to be very important for the recombination problem.}
\label{fig:Cic}
\end{figure}
\subsubsection{Importance of the different collisional processes}
\label{sec:importance_collisions}
To understand the effect of collisions and at which $n$ they are expected to become important, it is illustrative to compare the main collisional rates with the radiative rates in the CMB blackbody radiation field. Here in particular stimulated emission and photon excitation are important, and we included both into our computations.

In Fig.~\ref{fig:l-colls} we show the comparison of the $l$-changing collision rate with the total radiative rate, including emission, excitation and ionization processes, as a function of $l$ for different $n$. We computed the rates for the mixture of electrons and protons as obtained with the {\sc Recfast} code. At the considered redshifts helium is already completely neutral, and hence was excluded.
As one can see, at $z\sim 200$ only the very high Rydberg states are expected to be mixed over $l$ by collisions, while states with $n\lesssim 200$ should start to drop out of full statistical equilibrium\footnote{\changeREF{In full SE one has $N_{nl}=(2l+1)\, N_{n \rm s}$ for the level populations within a given shells $n$.}} (SE). 
At $z\sim 600$ one finds $n_{l-\rm mix}\sim 150$, and at $z\sim 1000$ even states with $n\sim 100$ should start to be completely mixed. 

This actually suggests another simplification of our recombination code. For levels with $n\gtrsim 300-400$ it should always be possible to treat them as if they are in full SE.
This should allow to add many more shells to the recombination problem, 
\changeREF{since it will be possible to obtain the solution for the populations within a given shell $n>n_{l-\rm mix}$ with only one additional differential equation per shell.
However,  at this stage we have not followed this possibility any further. 
}

We also checked the effect of collisional ionization. In Fig.~\ref{fig:Cic} we show the importance of collisional ionization by electron impact relative to the photoionization rate as a function of redshift.
Here it is important that according to the used approximations the collisional ionization rate is directly proportional to the photoionization cross sections. Therefore collisional ionization, like photoionization, is only effective from low-$l$ states ($l/l_{\rm max}\sim 0.1-0.3$). 
Even if collisional ionization can become comparable to the normal photoionization rate, in particular for the high $l$-states normal radiative bound-bound transitions will dictate the evolution of the levels. 
Collisional ionization therefore never seems to be very important for the recombination problem.
Our numerical computations also confirm this statement, where only the effect of $l$-changing collisions seems to be notable.

Regarding collisional excitations and de-excitations, we found that for $\alpha$-transitions ($\Delta n=1$) among excited levels, with the formulae given by \citet{Rege1962} at $z\sim 1100$ already for $n\sim 30-50$ these rates become comparable to the radiative transitions.
However, here it is important that for most of the levels in some given shell $n$ the $\alpha$-transitions are not defining the populations of the levels, so that the effect on the recombination process remains small up to much larger $n$.
Therefore, also collisional bound-bound transitions should be of minor importance for the recombination problem, but one will have to refine the modelling of these rates for detailed predictions of the low-frequency recombination spectrum ($\nu\lesssim 1\,$GHz). There energy changing collisional transitions and collisional ionizations are expected to suppress the emission of photons during the recombination epoch.
 We find such behaviour when artificially enhancing the collisional excitation and ionization rates.
 
 \changeREF{We would like to note, that collisional excitation and de-excitation, and collisional ionization at some very high $n$ are expected to push the populations of levels towards an equilibrium with the continuum. Since collisions are mediated by particle impact, the temperature of the electrons will be important for these levels. 
However, in our computations up to $n_{\rm max}=300$ the populations of the levels were always more close to Boltzmann equilibrium with the lower states, where the CMB blackbody temperature is important.
Nevertheless, even for those shells that started to be $l$-mixed in our computations the deviations from Boltzmann equilibrium with lower states were still rather large. 
All this suggests that for additional simplification of the recombination problem, the probably most straightforward approach will be to represent $l$-mixed shells with only one additional differential equations, but to account for the evolution with $n$ in detail.
}

\begin{figure}
\centering
\includegraphics[width=0.95\columnwidth]{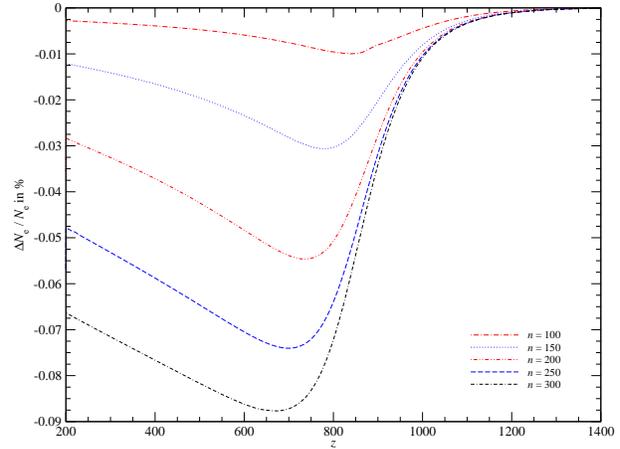}
\caption{Correction to the ionization history caused by collisional processes for different number of hydrogen shells as a function of redshift.}
\label{fig:Xe.colls}
\end{figure}
\subsubsection{Effect on the cosmological ionization history}
\label{sec:Xe_colls}
%
In Fig.~\ref{fig:Xe.colls} we present the correction to the ionization history caused by collisions. According to our computations $l$-changing collisions are dominating the corrections. 
At high redshifts, collisions do not affect the recombination history much. There the effective recombination rate is already saturated, so that mixing among the high $l$ states, which should lead to an increase in the effective recombination coefficient, does not result in any modification.
At low redshifts ($z\lesssim 1000$), the increase in the effective recombination coefficient mediated by $l$-changing collisions, leads to a small acceleration of recombination. 

\changeREF{At this point we did not treat the effect of collisions for more than 300 shells.
However,} one does expect some additional changes when going to a larger number of shells, as the fully mixed levels will continue to accelerate recombination. 
\changeREF{But}, here it will be very important to refine the calculations of the collisional rates, as with the current accuracy they may easily be up to a few times off. 
We therefore, stopped at this point and will return to this problem, once we have better estimates for the collisional rates.

\begin{figure}
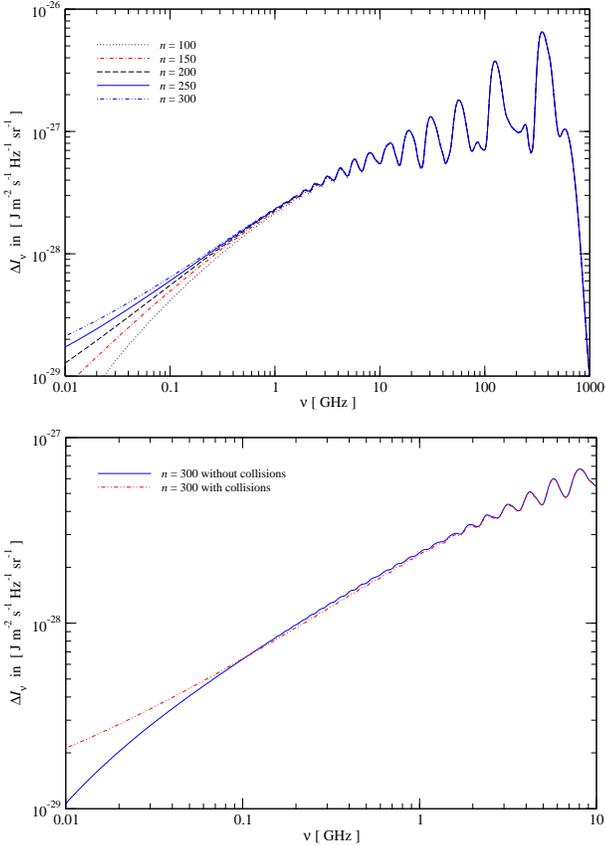

\centering
\includegraphics[width=0.95\columnwidth]{./eps/Spec.colls.eps}
\\[2mm]
\includegraphics[width=0.95\columnwidth]{./eps/Spec.colls.comp.eps}
\caption{Effect of collisions on the bound-bound cosmological recombination spectrum of hydrogen. For the upper panel $l$-changing collisions, collisional excitation and ionization were included. In the lower panel we show the direct comparison of the obtained spectra for $n_{\rm max}=300$ with and without collisions included. \changeII{Free-free absorption was not included.}}
\label{fig:spec.colls}
\end{figure}
\subsubsection{Effect on the cosmological recombination spectrum}
\label{sec:spec_colls}
In Fig.~\ref{fig:spec.colls} we show the effect of collisions on the cosmological recombination spectrum. In particular at low frequencies ($\nu\lesssim 1\,$GHz) collisions modify the recombination radiation in comparison to the case without collisions (see Fig.~\ref{fig:spec}). Due to $l$-changing collisions, more electrons reach the high $n$ and high $l$ states. From these levels only transitions with small $\Delta n$ are possible, so that more photons are emitted at low frequencies, explaining the enhancement there.
We would like to mention, that for the considered cases, $n$-changing collisions and collisional ionizations are still not very important. 

We also note that at this stage the uncertainty in the collisional rates does not allow to compute precise templates for the recombination spectrum at $\nu\lesssim 1\,$GHz.
Also in connection with the ionization history, the final modification caused by collisions cannot be given at this point. 
It will therefore be important to refine the treatment of collisional processes.
We plan to investigate this problem in the near future.

%
%
%
\section{Conclusions}
\label{sec:conc}
In this paper we compute the cosmological recombination history including a total of $\sim 61\,000$ states into the hydrogen atom.
For this purpose we developed a new ODE solver that allows us to treat large systems of very stiff differential equations.
This solver was implemented in a general way, so that it also can be applied to other problems, e.g. in computations of nuclear networks appearing in supernova and star formation calculations, or for chemical networks important during reionization and the dark ages.

We discussed the effect of recombinations to the high Rydberg states in hydrogen on the dynamics of recombination and the recombination spectrum. 
Our computations indicate that for the cosmological ionization history at redshifts $z\gtrsim 200$ the atomic model for hydrogen becomes complete for $n_{\rm max}\sim 300-350$.
At $z\sim 200$ we find a modification of $\Delta N_{\rm e}/N_{\rm e}\sim 1.6\%$ when including 350 shells.
This is about a factor of $1.8$ smaller than the result for a 100-shell computations (see Sect.~\ref{sec:Xe} and Fig.~\ref{fig:Xe} for more details).
Nevertheless, the effect on the CMB temperature and polarization power spectra at large multipoles is small when going from 100 to 350 shells (see Fig.~\ref{fig:DCl}).
From the point of view of {\sc Planck} data analysis $\sim 100$ shells seem to be already enough, and the conclusions reached by \citet{Jose2010} for the biases to $n_{\rm s}$ and $\Omega_{\rm b} h^2$ caused by the detailed physics of recombination should remain practically the same.
However, the total value of $\tau$ could still be affected at a significant level by the discussed process. 

Due to the huge improvements in the performance of our recombination code it is possible to compute the recombination history for 350 shells in about a day (see Table~\ref{tab:perform}).
\changeREF{Although according to our computations 100 shells seem to be sufficient for {\sc Planck} data analysis, we} still plan to provide an updated training set for {\sc Rico} \citep{Fendt2009}, which then could be used for \changeREF{precise} computations of the CMB power spectra.
Also for the final code comparison, this type of updated training set may be useful.
\changeREF{Alternatively, we plan to investigate possibilities for a refined treatment within a {\sc Recfast} type of scheme, as mentioned in Sect.~\ref{sec:RECFASTUPDATE}.}

The inclusion of more shells to hydrogen also leads to an increase of the amplitude of the cosmological recombination spectrum at low frequencies, since recombinations to highly excited states allow additional electrons to emit photons in transitions with small $\Delta n$.
The main results for the recombination spectrum are presented in Sect.~\ref{sec:spec} and in particular Fig.~\ref{fig:spec}.
At frequencies around and above $\sim 1\,$GHz, the recombination spectrum converges at the level of percent or better, when including 350 shells. 
Such precision will become important in the future when using measurements of the cosmological recombination spectrum to constrain  cosmological parameters \citep{Sunyaev2009}.
Nevertheless, additional non-standard processes may also be important here, e.g. due to pre-recombinational energy release \citep{Chluba2008c}, or annihilation of dark matter particles \citep{Chluba2010a}.

We also discussed the effect of collision on both the dynamics of recombination and the recombination spectrum (Sect.~\ref{sec:Xe_colls} and \ref{sec:spec_colls}).
Our analysis suggests that at $z\gtrsim 200$ levels with $n\gtrsim 300-400$ will always be completely mixed over $l$.
Collisional ionizations and excitations, however, seem to become important only for much larger $n$.
If states are completely mixed over $l$ it will, in principle, become possible to add many more shells to the recombination problem, since only one additional equation per shell will be required above $n_{\rm l-mix}$. 
We plan to investigate this possibility in the near future, however, at this point it will be more urgent to refine the computations of collisional rate coefficients. Here we only used very rough approximations, common for computations in stellar astrophysics.
However, both the cosmological recombination spectrum and the recombination dynamics require updated and more precise computations of these rates. 
We hope that this work will motivate some experts in atomic physics to attack this complicated problem in the near future. 

With our current estimates we find a correction to the free electron number density of $\Delta N_{\rm e}/N_{\rm e}\sim -\pot{8.8}{-4}$ at $z\sim 700$, which is mainly caused by $l$-changing collisions with protons.
However, this result could be off by factors of a few because of the uncertainties in the used collisional rates.
Also, for the final answer one will probably have to include more shells into the computation, as the convergence of this correction is very slow with $n_{\rm max}$. Our current computations with collisions were limited to $n_{\rm max}=300$, however, according to our estimates $n_{\rm max}\sim 300-400$ will likely be necessary. 
For computations of the CMB power spectra it will be important to get this correction right.

\section*{Acknowledgements}
\changeREF{The authors would like to thank the anonymous referee for useful comments and suggestions.}
JC is grateful to R.A.~Sunyaev for useful discussions and suggestions, \changeREF{and hospitality during his visit to MPA in Feb/March 2010}.
Furthermore, JC is very glad that he had a \changeREF{chance to talk to D.~Grin about {\sc RecSparse} during his visit to CITA in Nov. 2009.}
JC would also like to thank K.~Dolag and M.~Reinecke for useful discussions on numerical issues, \changeII{M.~Bergemann for stimulating discussions on collisional processes}, \changeREF{and Y.~Ali-Ha\"{i}moud for carefully reading the manuskript}.
%
Furthermore, the authors would like to acknowledge the use of computational resources at MPA.
\changeI{Also, \changeREF{several} computations were performed on the GPC supercomputer at the SciNet HPC Consortium. 
SciNet is funded by: the Canada Foundation for Innovation under the auspices of Compute Canada; the Government of Ontario; Ontario Research Fund - Research Excellence; and the University of Toronto.
}

\begin{appendix}

%
%
%
\section{Coefficients for the Gear's method with variable order and time-step}
\label{app:Gears_coeffies}
To obtain the coefficients for the variable time-step implicit Gear's formula Eq.~\eqref{eq:dy_dt_Gear} and the extrapolation formula Eq.~\eqref{eq:dy_dt_Extra} it is convenient to define 
\beal
\label{eq:Dt_defs}
\Delta t_{i-k}=t_{i-k}-t_i
\\
\rho_k=\frac{\Delta t_{i-k}}{\Delta t_{i+1}}.
\end{align}
Here $k=0, 1, 2, 3, 4$ when the solution at 4 previous and the current time $t_i$ are available. For $k=0$ one has $\Delta t_i=0$ and $\rho_k=0$.

\subsection{Coefficients for implicit Gear's formulae}
\label{app:Gears_form_coeffies}
Using the definitions given above one finds
\bsub
\label{eq:alpha_i}
\beal
\lambda&=1+\sum_{k=1}^5 \rho_k\,\kappa_k
\\
\kappa_j
&=\frac{1}{\rho_j f_{j}(\rho_j)}\,\left[(-1)^j\,\prod_{k=1}^{j-1} \frac{(1+\rho_k)^2}{\rho_j-\rho_k}
- \sum_{k=j+1}^5 \rho_k\,\kappa_k\,f_{j}(\rho_k)\,\prod_{m=1}^{j-1} \frac{\rho_k-\rho_m}{\rho_j-\rho_m}\right]
\end{align}
with
\beal
f_0(x)&=x^{-1}
\\
f_{1}(x)&=2+x
\\
f_{2}(x)&=f_{1}(\rho_1)\,f_{1}(x)-1
\\
f_{3}(x)&=f_{2}(\rho_2)\,f_{1}(x)-f_1(\rho_1+\rho_2)
\\
f_{4}(x)&=f_{3}(\rho_3)\,f_{1}(x)-f_1(\rho_1+\rho_2)\,\rho_3-f_{2}(\rho_2)
%
\\
f_{5}(x)&=f_{4}(\rho_4)\,f_{1}(x)-\left[ f_2(\rho_2+\rho_3) + \rho_2\,\rho_3 \right] \,\rho_4
-f_{3}(\rho_3)
\end{align}
\esub
One has to calculate the $\kappa_i$ starting with $\kappa_5$. To use a lower order Gear's formula $l<5$ one has to set $\kappa_i=0$ for $i>l$. Note that in our notation $\prod_k^j [...]\equiv 1$ for $k\geq j$.

\subsection{Coefficients for extrapolation}
\label{app:extrapol_coeffies}
Using the definitions given above one finds
\beal
\label{eq:beta_i}
\gamma_0&=1-\sum_{k=1}^5 \gamma_k
\\
\gamma_j
&=\frac{1}{\rho_j}\,\left[(-1)^j\,\prod_{k=1}^{j-1} \frac{1+\rho_k}{\rho_j-\rho_k}
- \sum_{k=j+1}^5 \rho_k\,\gamma_k\,\prod_{m=1}^{j-1} \frac{\rho_k-\rho_m}{\rho_j-\rho_m}\right]
\end{align}
for $j\leq 5$. One has to calculate the $\gamma_j$ starting with $\gamma_5$. To calculate the extrapolation based on $1<l<5$ previous time-steps  one has to set $\gamma_j$ for $j>l$ equal to zero. Note that again $\prod_k^j [...]\equiv 1$ for $k\geq j$.

\end{appendix}

\bibliographystyle{mn2e} 
\bibliography{Lit}

\end{document}